\documentstyle[emulateapj,flushrt]{article}
\newcommand{\beq}{\begin{equation}}
\newcommand{\eeq}{\end{equation}}
\newcommand{\rem}[1]{#1}\newcommand{\remm}[1]{ }
\begin{document}
\title{Interstellar Scintillations of Polarization of Compact Sources}

\author{Mikhail V. Medvedev\altaffilmark{1} }

\affil{Harvard-Smithsonian Center for Astrophysics, 60 Garden Street,
Cambridge, MA 02138}

\altaffiltext{1}{Also at the Institute for Nuclear Fusion, RRC ``Kurchatov
Institute'', Moscow 123182, Russia; E-mail: mmedvedev@cfa.harvard.edu; URL:
http://cfa-www.harvard.edu/\~{ }mmedvede/ }

\begin{abstract}
We demostrate that the measurement of fluctuations of polarization 
due to the galactic interstellar scintillations may be used 
to study the structure of the radiation field at compact radio sources.
We develop a mathematical formalism and demonstrate it on a simple analytical 
model in which the scale of the polarization variation through the source 
is comparable to the source size. The predicted amplitude of modulation of 
the polarized radiation flux is $\sim20\%\times\pi_{\rm s}\times m_{\rm sc}$, 
where $\pi_{\rm s}$ is the characteristic degree of polarization of radiation 
at the source and $m_{\rm sc}$ is the typical modulation index due to
scattering, i.e., $m_{\rm sc}\simeq1$ for diffractive scintillations and 
$m_{\rm sc}<1$ for refractive scintillations.
\end{abstract}

\section{Introduction}

Interstellar scintillations of compact radio sources, which were first
identified in the short-term variability of pulsars (\cite{Scheuer68}; 
\cite{Ricket69}) and explained as random scattering and diffraction from 
the electron density fluctuations in the ionized interstellar plasma,
have been used in studies of galactic and extra-galactic objects
(see, e.g., \cite{Cordes-etal85}; 1986), mostly pulsars, and also active 
galactic nuclei (see, e.g., \cite{Ghosh-etal90} and references therein),
some compact extra-galactic radio sources (e.g., \cite{Rickett-etal95}), 
$\gamma$-ray burst remnants (\cite{Goodman97}; \cite{Waxman-etal98}), 
and the accretion disk around the galactic center (\cite{GC}). 
However, this technique is (mostly) limited to the measurements of the 
intensity variations, while the information
contained in the polarization properties is usually lost.
In this letter we extend the interstellar scintillation observation 
technique to include the detection of polarization fluctuations, which
will allow to study the structure of the radiation field at the source. 
For synchrotron radiation, for instance, this allows to infer the geometry and 
correlation properties of the magnetic fields, --- the information which is 
difficult to obtain by other techniques because of resolution limits
(see, e.g., \cite{ML99} for an application to $\gamma$-ray bursts).

In this letter we develop a formalism of the interstellar scintillations of 
polarization. For simplicity, we do not consider frequency-related effects.
Using a simple model of the source radiation, we {\em analytically} calculate 
modulation indices. We demonstrate that the amplitude of fluctuations 
of polarization is maximum when the angular correlation scale of radiation at 
the source (which is often the source size) is comparable to the typical
scale of `speckles' produced at the observing plane by the interstellar
scattering, and amounts up to 
$\sim20\%\times\pi_{\rm s}\times m_{\rm sc}$ modulation, where $\pi_{\rm s}$ 
is the typical degree of polarization at the source and $m_{\rm sc}$ is the 
typical modulation amplitude which is $\simeq1$ for diffractive scintillations 
and usually $\sim0.1$ for refractive scintillations.

The rest of this letter is organized as follows. In \S \ref{S:FORMALISM}
we present a mathematical technique, which we
then demonstrate on a simple analytical model in \S \ref{S:MODEL}.
We present our conclusions in \S \ref{S:CONCL}.

\section{Polarization Scintillation Formalism  \label{S:FORMALISM} }

The interstellar scintillations arise when fluctuations in the electron density
randomly modulate the refractive index of a medium. As a result of diffraction 
(diffractive scintillations) and random focusing (refractive scintillations) of
electromagnetic waves, a point source produces a pattern in an observing plane 
which consists of randomly located 
bright and dim spots --- speckles --- and observed as fluctuations of 
the source brightness when an observer moves through the pattern. 
The characteristic correlation length of the pattern, $r_0(\lambda)$, is
determined by the regime of scattering, statistical properties of the 
interstellar turbulence, and the observing wavelength, $\lambda$ 
(for more details, see a nice review by \cite{Narayan92}).\footnote{
	Let us introduce the Fresnel scale, 
	$r_{\rm F}=\left(\lambda D/2\pi\right)^{1/2}$
	(where $D$ is the distance to the scattering screen) and the
	diffractive length, $r_{\rm diff}$, which represents the transverse 
	separation for which the root mean square phase difference (produced by
	scattering) is equal to one radian. In the weak scattering regime 
	($r_{\rm diff}\gg r_{\rm F}$), the speckle size is of order of the
	Fresnel scale, $r_0\sim r_{\rm F}$. In the strong scattering regime 
	($r_{\rm diff}\ll r_{\rm F}$), the diffractive and refractive effects
	are distinct (\cite{GN85}). For the short-term diffractive
	scintillations, the speckle size is $r_0\sim r_{\rm diff}$, while for
	the longer-term refractive scintillations, it is 
	$r_0\sim r_{\rm ref}=r_{\rm F}^2/r_{\rm diff}\gg r_{\rm diff}$. 
	Note that the apparent angular size of the scattered point
	source image is $\theta_{\rm scatt}\sim r_{\rm ref}/D$ for both 
	diffractive and refractive scintillations. \label{foot} }
It is convenient
to introduce the `angular' size of a speckle, $\theta_0=r_0/D$, where $D$ is
the distance to the scattering screen, and we define $\theta_{\rm s}$ to be
the angular size of a source. 

Let us assume that the emission is polarized, and the sense of polarization 
varies through the source, so that the net (average) 
polarization vanishes.\footnote{
	If the net degree of polarization 
	is nonzero, one can always subtract the constant contribution.
	The variation of this polarized contribution is identical to the 
	total flux variation. }
If $\theta_{\rm s}\ll\theta_0$, the source is effectively point-like. In this 
regime the intensity variations are large, while the 
polarization signal is weak due to averaging. In contrast, when 
$\theta_{\rm s}\gtrsim\theta_0$, different elements in the source produce
identical patterns, which are offset by $D\,\delta\theta$, according to the 
element angular separation, $\delta\theta$.
The overall pattern is, thus, obtained from the superposition
(convolution) of individual, incoherent patterns. An observer 
will measure fluctuations of the direction of polarization, while 
the intensity exhibits less modulation. When $\theta_{\rm s}\gg\theta_0$, 
the offsets become much larger than the speckle size, so that the speckles 
are multiply overlapped and both intensity and polarization fluctuations 
smear out.

The polarization properties of radiation are determined by the second-rank
polarization tensor. The correlation properties of the scintillation
pattern are therefore described by a fourth-rank correlation matrix. If,
however, the scattering medium is `passive,' i.e., does not affect the
polarization, then most of the information
contained in this correlation matrix is redundant. The properties of radiation 
are then fully described by four scalar quantities, --- the Stokes
parameters, which are additive for independent (incoherent) radiation 
fluxes (\cite{RybickiLightman}). 
They are the intensity $I$ and $Q,\ U$, and $V$.
The quantities $Q$ and $U$ describe linear polarization, and 
$V$ describes circular polarization. 
The degree of polarization is defined as 
\beq
\pi=I_{\rm pol}/I=\left(Q^2+U^2+V^2\right)^{1/2}/I,
\label{degree-def}
\eeq 
where $I_{\rm pol}$ is the polarized intensity. Note that the polarized 
and total intensities are proportional to the integrated fluxes from the 
source which are directly measurable.

We assume that scintillations do not affect polarization properties 
of radiation. This is valid if either the projected turbulence is (on average) 
isotropic, i.e., there is no preferred direction in space, 
or the geometrical size of the inhomogeneities which contribute to
scattering is much larger than a wavelength, $s\gg\lambda$. Indeed, if 
$s\sim\lambda$ and there is a preferred direction in space, then 
transmission coefficients through a scintillation screen (`diffraction grid') 
of waves with different orientations of the plane of polarization with respect 
to that preferred direction will be different. 
It has been demonstrated (\cite{Simonetti-etal84}) that
birefringence in the interstellar medium is negligible in discussion of
scattering. Also, the effect of angular Faraday depolarization across a
source is weak for the observing frequencies above $\sim10\textrm{ MHz}$
(\cite{Linfield96}).

Let us assume that a point source is located at infinity. It illuminates a thin 
phase changing screen located at a distance $D$ from an observer plane $(x,y)$
and produces a speckle pattern on this plane. The statistics of speckles 
is determined by the screen (i.e., the interstellar turbulence), 
and is usually characterized by the second moment 
correlator of the complex electric field of the electromagnetic wave,
$\langle E({\bf r})E^*({\bf r}+\Delta{\bf r})\rangle
=\exp\left[-D_\phi(\Delta{\bf r})/2\right]
\propto\exp\left[-(\left|\Delta{\bf r}\right|/r_{\rm diff})^{\beta-2}\right]$,
where ${\bf r}=(x,y)$ and $\Delta{\bf r}$ are two-dimensional vectors, 
$\langle\ldots\rangle=\int\!\!\int\ldots{\rm d}x{\rm d}y$ denotes a 
spatial average, the `star' denotes complex conjugation, 
and $\beta$ is the power-law index of the power 
spectrum of the electron density fluctuations, 
$|\delta n_e(q)|^2\propto q^{-\beta}$ with $q$ being the spatial wave-number.
The quantity $D_\phi$ is the phase structure function which yields
the root mean square phase shift along different paths.
The value of $\beta$ for the galactic interstellar medium is 
somewhat uncertain but close to the Kolmogorov theory prediction $\beta=11/3$ 
(\cite{Armstrong-etal95}). In calculating the modulation indices below, 
we adopt the approximate value of $\beta\approx 4$ for which the correlator
is Gaussian, which greatly simplifies the calculation.

The intensity distribution, $I_{\rm p}(x,y)$, in the speckle pattern produced 
in observer's plane by a point source is determined by the Fourier 
transform of the electric field correlator. We introduce the normalized 
intensity distribution in the pattern and its Fourier transform 
as follows,
\beq
W(x,y)=\tilde I_{\rm p}/\langle I_{\rm p}\rangle
\equiv\left(I_{\rm p}-\langle I_{\rm p}\rangle\right)/\langle I_{\rm p}\rangle
\rightleftharpoons{\cal W}(k_x,k_y), 
\eeq 
where $\rightleftharpoons$ denotes a Fourier conjugated pair and the
`calligraphic' letters are used to denote the Fourier transformed quantities.
Since the Stokes parameters are additive, the point source speckle patterns
of $Q_{\rm p}$, $U_{\rm p}$, and $V_{\rm p}$ are also described by $W$. 

The intensity distribution produced by an extended source is 
obtained from the convolution of the intensity distribution 
of a point source with the source surface 
brightness distribution, $P_I(\theta_x,\theta_y)$,
\begin{eqnarray}
\tilde I(x,y)&=&W(x,y)\ast P_I(\theta_x,\theta_y) 
\nonumber\\
&\equiv&\int\!\!\!\int
W(x-D\theta_x,y-D\theta_y)\,P_I(\theta_x,\theta_y)\, 
{\rm d} \theta_x\, {\rm d} \theta_y 
\nonumber\\
&\rightleftharpoons&{\cal W}(k_x,k_y)\,{\cal P}_I(Dk_x,Dk_y)
=\tilde{\cal I}(k_x,k_y),
\label{I}
\end{eqnarray}
where %$D$ is the distance from an observer to the scattering screen, 
$(\theta_x,\theta_y)$ are angular coordinates in the sky, and we
normalize the total flux to unity, i.e., 
$\langle I\rangle=\int\!\!\int P_I(\theta_x,\theta_y)
{\rm d} \theta_x\, {\rm d} \theta_y=1$.
In equation (\ref{I}), the convolution theorem have been used.
The amplitude of intensity fluctuations due to scintillations is
determined by the modulation index:
\beq
m_I=\langle\tilde I^2\rangle^{1/2}
=\left(\int\!\!\!\int\left|\tilde{\cal I}(k_x,k_y)\right|^2\,
{\rm d}k_x\,{\rm d}k_y\right)^{1/2},
\label{Si}
\eeq
where the last expression follows from Parseval's theorem. 

By analogy with intensity, we define the distributions 
and modulation indices of the other Stokes parameters,
$$
\tilde Q=W\ast P_Q, \quad 
\tilde U=W\ast P_U, \quad 
\tilde V=W\ast P_V ,
\eqno{(\textrm{\ref{I}}')}
$$
$$
m_Q=\langle\tilde Q^2\rangle^{1/2}, \quad
m_U=\langle\tilde U^2\rangle^{1/2}, \quad
m_V=\langle\tilde V^2\rangle^{1/2} .
\eqno{(\textrm{\ref{Si}}')}
$$
The amplitude of fluctuations of the polarized intensity is described by 
\beq
m_{QUV}=\left(m_Q^2+m_U^2+m_V^2\right)^{1/2}.
\label{Squv}
\eeq 
By analogy with equation (\ref{degree-def}), we formally introduce 
a measure of the `degree' of polarization, $m_\pi=m_{QUV}/m_I$.

\section{Analytical Model \label{S:MODEL} }

An analytical calculation of the modulation indices is a difficult
mathematical task, even for relatively simple models of source emission and
interstellar turbulence. For this reason, we consider the simplest source 
model. We assume that the source has a uniform surface brightness over a
rectangular angular region on the sky: 
$-\theta_{\rm s}\le\theta_x\le\theta_{\rm s}$,\ 
$-\theta_{\rm s}\le\theta_y\le\theta_{\rm s}$,\ with $\theta_{\rm s}$ being the 
angular size (`radius') of the source. This approximation simplifies 
considerably the related integrals and provides a good approximation to a 
circular source. We normalize the total flux to unity. We choose the 
surface distribution of the Stokes parameters to be,
\begin{mathletters}
\begin{eqnarray}
P_I(\theta_x,\theta_y)&=&\frac{1}{4\,\theta_{\rm s}^2} 
\left[\Theta(\theta_x+\theta_{\rm s})-\Theta(\theta_x-\theta_{\rm s})\right] 
\nonumber\\ & &{ }\qquad
\times\left[\Theta(\theta_y+\theta_{\rm s})-\Theta(\theta_y-\theta_{\rm s})
\right] , \\
P_Q(\theta_x,\theta_y)&=&\pi_{\rm s}\,P_I(\theta_x,\theta_y)\,\frac{1}{2}
\left(\cos\frac{\pi \theta_y}{\theta_{\rm s}}
-\cos\frac{\pi \theta_x}{\theta_{\rm s}}\right), \\ 
P_U(\theta_x,\theta_y)&=&\pi_{\rm s}\,P_I(\theta_x,\theta_y)\,
\sin\frac{\pi\,\theta_x}{2\,\theta_{\rm s}}\,
\sin\frac{\pi\,\theta_y}{2\,\theta_{\rm s}}, \\
P_V(\theta_x,\theta_y)&=&0,
\end{eqnarray}
\label{modelP}
\end{mathletters}
\noindent
\noindent
where $\pi_{\rm s}$ is the `intrinsic' degree of polarization of 
radiation and $\Theta(x)$ is the Heaviside step-function. The functions $P_Q$ 
and $P_U$ are shown in Figure \ref{f:QU}. These distributions mimic the 
radiation field which is linearly polarized in the radial direction from 
the center of the source. We take the intensity distribution function to be 
Gaussian,
\beq
W(x,y)=\frac{\sqrt{2}\, m_{\rm sc}}{\sqrt{\pi} D\theta_0}
\exp\left[-\frac{\left(x^2+y^2\right)}{D^2\theta_0^2}\right],
\label{modelW}
\eeq 
which is normalized so that $\langle W^2\rangle^{1/2}=m_{\rm sc}$, where 
$m_{\rm sc}$ is the `typical' modulation index for a point source.
We do not specify $\pi_{\rm s}$, $\theta_0$, and $m_{\rm sc}$ here and keep
them as free parameters. The degree of polarization, $\pi_{\rm s}$, depends on
the emission mechanism and may be as high as $\sim 70\%$ for synchrotron 
radiation. The quantities $\theta_0$ and $m_{\rm sc}$ depend on the regime of 
scintillations and the wavelength of radiation. Namely, 
$\theta_0\sim r_{\rm diff}/D$ and $m_{\rm sc}\simeq1$ for diffractive 
scintillations, and $\theta_0\sim r_{\rm ref}/D$ and $m_{\rm sc}<1$ for 
refractive scintillations (see \cite{Narayan92} and also footnote \ref{foot}).

Using the formalism from the previous section (\S \ref{S:FORMALISM}),
we calculate the modulation indices $m_I$,\ $m_{QUV}$, and $m_\pi$.  Upon
lengthy and cumbersome mathematics, some details of which are presented in
Appendix, we arrive at the following expressions,
\begin{mathletters}
\begin{eqnarray}
m_I^2&=&\frac{m_{\rm sc}^2}{\pi^3\vartheta^2}\,{\cal A}(\vartheta)^2 ,\\
m_Q^2&=&\frac{\pi_{\rm s}^2 m_{\rm sc}^2}{4\,\pi^3\vartheta^2}
\left[\left[{\cal B}(\vartheta)+{\cal C}(\vartheta)\right]\,{\cal A}(\vartheta)
-2\,{\cal C}(\vartheta)^2\right] ,\\
m_U^2&=&\frac{\pi_{\rm s}^2 m_{\rm sc}^2}{4\,\pi^3\vartheta^2}
\left[{\cal B}(\vartheta)-{\cal C}(\vartheta)\right]^2, 
\end{eqnarray}
\label{result}
\end{mathletters}
where $\vartheta=\sqrt{2}\theta_{\rm s}/\theta_0$ is the size
of the object in units of the characteristic speckle scale, and 
${\cal A}(\vartheta),\ {\cal B}(\vartheta)$, and ${\cal C}(\vartheta)$ are
complicated expressions involving the error functions of real and imaginary
arguments, given by equations\ (\ref{A})--(\ref{C}). Using
equations\ (\ref{Si}),(\ref{Squv}), it is now straightforward to calculate
the quantities which are directly measured, namely the amplitude of
fluctuations of total and polarized intensity (flux), $m_I$ and $m_{QUV}$. The 
dependence of the scintillation indices on $\vartheta$ is presented 
in Figure\ \ref{scint}. Clearly, for a small source size,
($\theta_{\rm s}\ll\theta_0$), there are no fluctuations of 
polarization, while the intensity fluctuations are maximum 
$\sim 100\%\times m_{\rm sc}$. As the speckle size approaches the source size, 
the observed radiation is partially polarized. At the same time, the 
intensity contrast decreases due to the overlap between the speckles. 
The fluctuations of polarization peak when 
$\theta_{\rm s}\sim{\rm few}\times\theta_0$ at the value of 
$\sim20\%\times\pi_{\rm s}\times m_{\rm sc}$. 
For a larger source, the fluctuation amplitude of both polarized and
total intensity decreases. However, the `degree' of polarization 
($m_\pi=m_{QUV}/m_{I}$) continuously increases with the increasing source size 
and asymptotes at $\sim70\%\times\pi_{\rm s}$ in our model. 
Thus, the asymptotic value of $m_\pi$ is nearly independent of the scatterer 
and yields the intrinsic degree of polarization at the source within a factor 
of order unity.

\section{Conclusion \label{S:CONCL}}

In this paper we propose a new observational technique --- the detection of
polarization scintillations --- to study the structure of the radiation
field at compact radio sources. We have developed a general formalism and 
demonstrated it on a simple analytical model.
For randomly polarized radiation with the degree of polarization at the 
source typical of synchrotron radiation $\sim75\%$, the amplitude of 
fluctuations of the polarization flux is calculated to be $\sim15\%$ 
in the case of diffractive interstellar scintillations and an order 
of magnitude lower in the case of refractive scintillations.

We should note that these quantitative results were obtained for an 
over-simplified model and provide an illustration of the predicted effect.
Radiation fields of real sources are more complicated and are determined by the 
properties of emission processes and radiative transfer. The statistical 
properties of the interstellar turbulence are determined by pumping and
cascade mechanisms and the assumption that the fluctiations are gaussian
is also rather crude. Therefore, we expect that numerical values of the 
modulation indices computed in a more detailed study may be different from 
those presented here.

\acknowledgements 

The author is grateful to Ramesh Narayan interesting and useful discussions and 
his comments on the manuscript and Avi Loeb for interesting discussions.
This work was supported by NASA grant NAG~5-2837.

\begin{appendix}

\section{Calculation of the Scintillation Indices \label{A2} }

First, we introduce new dimensionless variables, $\xi=x/D,\ \eta=y/D$.
We use the following definition of the Fourier transform of $W$:
\beq
{\cal W}(k_\xi,k_\eta)=\int\!\!\!\int W(\xi,\eta)\,\exp
\left[-2\pi i(k_\xi\xi+k_\eta\eta)\right]\,{\rm d}\xi\,{\rm d}\eta,
\eeq
%\begin{mathletters}
%\begin{eqnarray}
%W(\xi,\eta)&=&\int\!\!\!\int {\cal W}(k_\xi,k_\eta)\,\exp
%\left[2\pi i(k_\xi\xi+k_\eta\eta)\right]\,{\rm d}k_\xi\,{\rm d}k_\eta, \\
%{\cal W}(k_\xi,k_\eta)&=&\int\!\!\!\int W(\xi,\eta)\,\exp
%\left[-2\pi i(k_\xi\xi+k_\eta\eta)\right]\,{\rm d}\xi\,{\rm d}\eta,
%\end{eqnarray}
%\end{mathletters}
and similarly for other quantities.
The Fourier transformed point source speckle distribution, 
equation\ (\ref{modelW}), is
\beq
{\cal W}(k_\xi,k_\eta)
=\sqrt{2\pi}\,m_{\rm sc}\,\theta_0\,
\exp\left[-\pi^2\theta_0^2\left(k_\xi^2+k_\eta^2\right)\right].
\eeq
The Fourier transformed distributions of intensity and the Stokes parameters,
equations\ (\ref{modelP}), expressed in terms of the new variables, 
are calculated to yield
\begin{mathletters}
\beq
{\cal P}_I(k_\xi,k_\eta)=
\frac{\sin 2\pi k_\xi\theta_{\rm s}}{\pi k_\xi}\,
\frac{\sin 2\pi k_\eta\theta_{\rm s}}{\pi k_\eta} ,
\eeq
\rem{  % one-line equation
\beq
{\cal P}_Q(k_\xi,k_\eta)=\pi_{\rm s}\,\theta_{\rm s}\left[
\frac{\sin 2\pi k_\xi\theta_{\rm s}}{2 \pi k_\xi}
\left(\frac{\sin \pi(2k_\eta\theta_{\rm s}-1)}{\pi(2k_\eta\theta_{\rm s}-1)}
+\frac{\sin \pi(2k_\eta\theta_{\rm s}+1)}{\pi(2k_\eta\theta_{\rm s}+1)}\right)
-\frac{\sin 2\pi k_\eta\theta_{\rm s}}{2 \pi k_\eta} 
\left(\frac{\sin \pi(2k_\xi\theta_{\rm s}-1)}{\pi(2k_\xi\theta_{\rm s}-1)}
+\frac{\sin \pi(2k_\xi\theta_{\rm s}+1)}{\pi(2k_\xi\theta_{\rm s}+1)}\right)
\right],
\eeq
}
\remm{  % two-lines equation
\begin{eqnarray}
{\cal P}_Q(k_\xi,k_\eta)&=&\pi_{\rm s}\,\theta_{\rm s}\Biggl[
\frac{\sin 2\pi k_\xi\theta_{\rm s}}{2 \pi k_\xi}
\left(\frac{\sin \pi(2k_\eta\theta_{\rm s}-1)}{\pi(2k_\eta\theta_{\rm s}-1)}
+\frac{\sin \pi(2k_\eta\theta_{\rm s}+1)}{\pi(2k_\eta\theta_{\rm s}+1)}\right)
\nonumber\\
& &{ }\qquad\qquad
-\frac{\sin 2\pi k_\eta\theta_{\rm s}}{2 \pi k_\eta} 
\left(\frac{\sin \pi(2k_\xi\theta_{\rm s}-1)}{\pi(2k_\xi\theta_{\rm s}-1)}
+\frac{\sin \pi(2k_\xi\theta_{\rm s}+1)}{\pi(2k_\xi\theta_{\rm s}+1)}\right)
\Biggr],
\end{eqnarray}
}
\beq
{\cal P}_U(k_\xi,k_\eta)=-\pi_{\rm s}\,\theta_{\rm s}^2
\left(\frac{\sin \pi(2k_\xi\theta_{\rm s}-1)}{\pi(2k_\xi\theta_{\rm s}-1)}
-\frac{\sin \pi(2k_\xi\theta_{\rm s}+1)}{\pi(2k_\xi\theta_{\rm s}+1)}\right)
\left(\frac{\sin \pi(2k_\eta\theta_{\rm s}-1)}{\pi(2k_\eta\theta_{\rm s}-1)}
-\frac{\sin \pi(2k_\eta\theta_{\rm s}+1)}{\pi(2k_\eta\theta_{\rm s}+1)}\right) .
\eeq
\end{mathletters}
The amplitude of fluctuations, which is the auto-correlation function at
zero separation, may now be calculated using Parseval's theorem [see
equation~(\ref{Si})]: 
\beq 
\langle\tilde I^2\rangle= \Bigl.\langle
\tilde I(\xi,\eta)\,\tilde I(\xi+\delta\xi,\eta+\delta\eta)\rangle
\Bigr|_{\delta\xi=0\atop\delta\eta=0} 
=\int\!\!\!\int{\cal W}(k_\xi,k_\eta)\,{\cal W}(-k_\xi,-k_\eta)\, 
{\cal P}_I(k_\xi,k_\eta)\,
{\cal P}_I(-k_\xi,-k_\eta)\,{\rm d}k_\xi\,{\rm d}k_\eta 
\eeq 
and similarly for $\langle\tilde Q^2\rangle$ and $\langle\tilde U^2\rangle$. 
In calculating these quantities, one encounters the following basic integrals: 
\beq 
I_1=\int_{-\infty}^{\infty}e^{-a^2\xi^2}\frac{\sin^2\pi \xi}{\xi^2}\,
{\rm d}\xi , 
\quad
I_2=\int_{-\infty}^{\infty}e^{-a^2\xi^2}\frac{\sin^2\pi(\xi-1)}{(\xi-1)^2}\,
{\rm d}\xi,
\quad
I_3=\int_{-\infty}^{\infty}e^{-a^2\xi^2}\frac{\sin^2\pi(\xi-1)}{\xi-1}\,
{\rm d}\xi.  
\eeq 
Other integrals can be reduce to the above integrals by noticing
that $\sin\pi(\xi-1)=\sin\pi(\xi+1)=-\sin\pi \xi$.  To illustrate the method 
of taking these integrals, consider $I_2$ as an example. We write $I_2$ as a
function of a new parameter, say $\alpha$, so that it reduces to the
original integral for $\alpha=1$, 
\beq
I_2(\alpha)=\int_{-\infty}^{\infty}e^{-a^2\xi^2}
\frac{\sin^2\alpha\pi(\xi-1)}{(\xi-1)^2}\,{\rm d}\xi .  
\eeq 
Differentiating twice with respect to this new parameter, we obtain an 
integral which can now be evaluated, 
\beq 
{I_2}''_{\alpha\alpha}(\alpha)
=2\pi^2\int_{-\infty}^{\infty}e^{-a^2\xi^2}\cos2\pi\alpha(\xi-1)\,{\rm d}\xi
=2\pi^{2}\,\frac{\sqrt{\pi}}{a}\,\exp\!\left(\frac{-\pi^2\alpha^2}{a^2}\right)
\cos2\pi\alpha.  
\eeq 
One should now integrate ${I_2}''_{\alpha\alpha}(\alpha)$ twice with respect 
to $\alpha$ and then set $\alpha=1$, 
\beq 
I_2=\int_0^1\left(\int_0^\alpha\,
{I_2}''_{\alpha\alpha}(\alpha_1)\,{\rm d}\alpha_1\right){\rm d}\alpha .
\eeq 
The lower limits in the integrals are taken so as to automatically
include constants of integration by taking into account that
${I_2}(0)={I_2}'_{\alpha}(0)=0$, by definition. The integrals involving the
error functions, which emerge after the first integration, may be taken by
integrating by parts and using the fact that 
${\rm erf}'(\xi)=(2/\sqrt{\pi})\,e^{-\xi^2}$. 
Finally, the integrals $I_1$--$I_3$ become, respectively, 
\beq 
{\cal A}(\vartheta)=\frac{\pi^{3/2}}{\vartheta}\,\left(e^{-\vartheta^2}-1
+\sqrt{\pi }\,\vartheta\,{\rm erf}(\vartheta)\right) ,
\label{A}
\eeq
\rem{  % one-line equation 
\beq
{\cal B}(\vartheta)=\frac{\pi^{3/2}}{\vartheta}\left[e^{-\vartheta^2}-1
-\frac{\sqrt{\pi}}{2}e^{-\pi^2/\vartheta^2}
\left[
\left(\frac{\pi}{\vartheta}-i\,\vartheta\right)
{\rm erfi}\!\left(\frac{\pi}{\vartheta}-i\,\vartheta\right)
+\left(\frac{\pi}{\vartheta}+i\,\vartheta\right)
{\rm erfi}\!\left(\frac{\pi}{\vartheta}+i\,\vartheta\right)
-2\,\frac{\pi}{\vartheta}\,{\rm erfi}\!\left(\frac{\pi}{\vartheta}\right)
\right]\right] ,
\label{B}
\eeq
}
\remm{  % two-line equation 
\begin{eqnarray}
{\cal B}(\vartheta)&=&\frac{\pi^{3/2}}{\vartheta}\Biggl[e^{-\vartheta^2}-1
-\frac{\sqrt{\pi}}{2}e^{-\pi^2/\vartheta^2}
\biggl[
\left(\frac{\pi}{\vartheta}-i\,\vartheta\right)
{\rm erfi}\!\left(\frac{\pi}{\vartheta}-i\,\vartheta\right)
\nonumber\\
& &{ }\qquad\qquad\qquad
+\left(\frac{\pi}{\vartheta}+i\,\vartheta\right)
{\rm erfi}\!\left(\frac{\pi}{\vartheta}+i\,\vartheta\right)
-2\,\frac{\pi}{\vartheta}\,{\rm erfi}\!\left(\frac{\pi}{\vartheta}\right)
\biggl]\Biggr] ,
\label{B}
\end{eqnarray}
}
\beq
{\cal C}(\vartheta)=\frac{\pi}{4}\,e^{-\pi^2/\vartheta^2}
\left[
{\rm erfi}\!\left(\frac{\pi}{\vartheta}-i\,\vartheta\right)
+{\rm erfi}\!\left(\frac{\pi}{\vartheta}+i\,\vartheta\right)
-2\,{\rm erfi}\!\left(\frac{\pi}{\vartheta}\right)
\right] ,
\label{C}
\eeq 
where ${\rm erf}(x)=2/\sqrt{\pi}\int_0^xe^{-t^2}{\rm d}t$ is the error
function, ${\rm erfi}(z)=2/\sqrt{\pi}\int_0^ze^{t^2}{\rm d}t=-i\,{\rm
erf}(i\,z)$ is the error function of imaginary argument, and
$\vartheta=\sqrt2\theta_{\rm s}/\theta_0$. The scintillation indices are then
readily calculated to yield equations (\ref{result}).

\end{appendix}

\figcaption[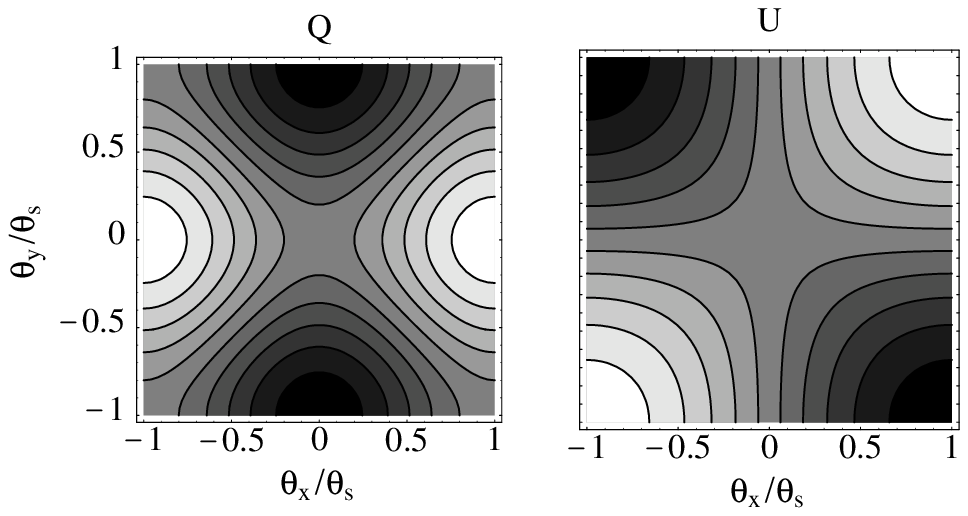]{Distributions of the Stokes parameters, $P_Q$ and
$P_U$, over the source in the analytical model. From black to white, 
$P$ varies from $-1$ to $+1$. \label{f:QU} }

\figcaption[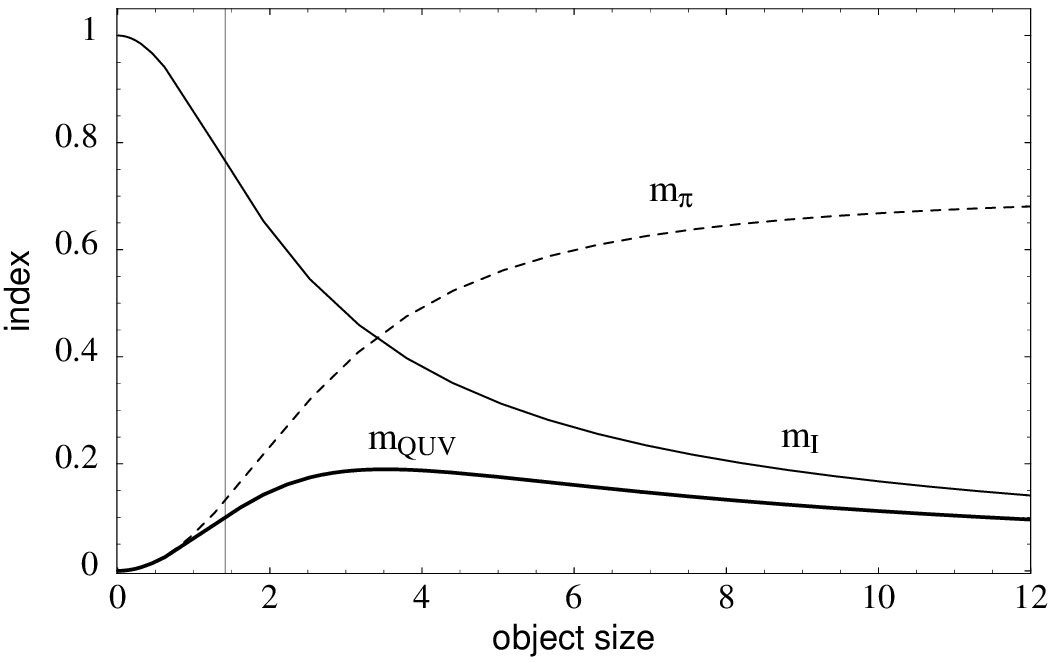]{Modulation indices of the intensity ($m_I$), 
polarized intensity ($m_{QUV}$), and `degree' of 
polarization ($m_\pi=m_{QUV}/m_I$), as functions of the normalized source 
size $\vartheta$ for completely polarized radiation, $\pi_{\rm s}=1$, in the
diffractive scattering regime, $m_{\rm sc}=1$.
The thin vertical line marks where $\theta_{\rm s}=\theta_0$. \label{scint} }

\vfil

\rem{
%\begin{figure}
\plottwo{dist-qu.eps}{scint.eps}\\
Figs.~\ref{f:QU},\ref{scint}
%\end{figure}
}

\remm{
\newpage
\plotone{dist-qu.eps}\\ \vskip0cm Fig.\ \ref{f:QU}\\ \vskip1cm
\plotone{scint.eps}\\ \vskip0cm Fig.\ \ref{scint}
}

\end{document}